\documentclass[a4paper,twocolumn,11pt,accepted=2024-02-28]{quantumarticle}

\pdfoutput=1
\usepackage[utf8]{inputenc}
\usepackage[english]{babel}
\usepackage[T1]{fontenc}
\usepackage{amsmath}
\usepackage{hyperref}
\usepackage[numbers]{natbib}

\usepackage{epsfig}
\usepackage{bbold}
\usepackage{epigraph}
\usepackage{relsize}
\usepackage{leftindex}

\newcommand{\ket}[1]{\left | #1 \right \rangle}
\newcommand{\bra}[1]{\left \langle #1 \right |}
\newcommand{\braL}[2]{\leftindex_{#1}{\bra{#2}}}

\newcommand{\beq}{\begin{equation}}
\newcommand{\eeq}{\end{equation}}

\begin{document}
\title{Teleportation of Post-Selected Quantum States}

\author{Daniel Collins}
\orcid{0009-0009-0365-4206}
\affiliation{H. H. Wills Physics Laboratory, University of
Bristol, Tyndall Avenue, Bristol BS8 1TL}

\begin{abstract}
Teleportation allows Alice to send a pre-prepared quantum state to Bob using only pre-shared entanglement and classical communication.
Here we show that it is possible to teleport a state which is also $\it{post}$-selected.
Post-selection of a state $\Phi$ means that after Alice has finished her experiment she performs a measurement and only keeps runs of the experiment where the measurement outcome is $\Phi$.
We also demonstrate pre and post-selected $\it{port}$-based teleportation, and use it as a quantum memory for such states.  Finally we use these protocols to perform instantaneous non-local quantum computation on pre and post-selected systems, and significantly reduce the entanglement required to instantaneously measure an arbitrary non-local variable of spatially separated pre and post-selected systems.

\end{abstract}

\maketitle

\section{Introduction}

\epigraph{Learn to see things backwards, inside out, and upside down}{\textit{The Tao of Leadership \\ John Heider}}

Teleportation is one of the most surprising and fundamental protocols of quantum mechanics.  In classical mechanics Alice can send a state to Bob without having any prior knowledge of it by measuring it as accurately as desired, sending the measurement results in a classical message to Bob, and having Bob recreate the state.  However in quantum mechanics the uncertainty relation limits the accuracy to which one can simultaneously measure all components of the state.  Thus it appears as if the only way to transfer the state from Alice to Bob is by transferring a particle explicitly carrying the state, e.g. by a fibre optic cable which sends polarized photons from Alice to Bob, and which maintains coherence between different polarizations for each photon.  To great surprise it was shown by Bennett et al. \cite{teleportation} that a state can be faithfully transferred from Alice to Bob without transferring any quantum particles, by using pre-shared entanglement, local measurements, classical communication and local unitary operations.  This has been experimentally performed many times starting with \cite{teleportationExperiment1, teleportationExperiment2} (see \cite{teleportReview} for a review).  Teleportation is now used as a fundamental building block for many tasks in quantum information and computation.

Usually when we talk about states in quantum mechanics we are thinking of states which are prepared in advance, i.e. pre-selected.  However Aharonov et al. \cite{postSelect,postSelectMore} showed that quantum mechanics also allows a post-selection of the state, which behaves quite differently to any post-selection in classical mechanics.  Post-selection means that we run an experiment on a system which ends with some sort of measurement, and only look at runs of the experiment where the measurement gives a particular outcome.  This post-selection is quite natural, e.g. we send photons through some optical circuit and only keep results where one of the final detectors receives a photon, we may select one polarization more than another through Polarization Dependent Loss \cite{telecomPostSelection}.  However it gives us information about the system at intermediate times which is impossible with pre-selection alone.  For example if we pre-select a particle with a well defined position and post-select it with a well defined momentum we will be able to predict the outcome of a position or momentum measurement of the particle at intermediate times with arbitrary accuracy, i.e. better than the uncertainly principle allows for pre-selected states.  This is quite different to deterministic classical mechanics, where all the information about a system can be pre-selected, and a post-selection is equivalent to a pre-selection. 

Having post-selected on a particular state, we can then evolve this final state backwards in time, and use it to retrodict the outcomes of measurements at earlier times.  We don't have to do this: we can make the same predictions using the usual pre-selected states, evolving the system forwards in time and conditioning on the final measurement.  However we find this time reversed method of thinking useful, and believe it allows us to more easily find interesting physics.

Post-Selection is a key part of the single photon source used for many quantum optic experiments \cite{parametricDownConversion}, internally creating a pair of photons via spontaneous parametric down-conversion, then post-selecting on one photon from the pair being detected, leaving the other as a single photon.  Combined with weak measurements \cite{weakValues}, post-selected states led to a method for improving the precision of certain measuring devices \cite{weakValueUses} which was used in the first observation of the spin Hall effect of light \cite{spinHallEffect}, and to measure the linear travel of a piezoelectric actuator down to the diameter of a Uranium nucleus \cite{ultrasensitiveMeasurement}.  Process matrices are equivalent to certain quantum systems between definite pre and post-selected states \cite{processMatrices}.  Post-selected states are also used for the intuition behind several curious paradoxical effects, e.g. the quantum pigeonhole principle \cite{pidgeonholePrinciple} and the quantum cheshire cat \cite{cheshireCat}.  

In this paper we describe teleportation of a post-selected state.  It was previously known \cite{nonLocalPreAndPostSelectedLev} that this could be accomplished for a $2$ dimensional system with probability $1/4$.  However like the original teleportation, we shall show that this can be done with certainty.  We then extend this to pre and post-selected states.  This is a fundamental protocol, which can then be used as a building block for more complex tasks.  We also extend $\it{port}$-based teleportation to pre and post-selected states, show how this can be used as a quantum memory for such states, and discuss instantaneous measurements and computation of pre and post-selected states.

Note that if teleportation was a unitary operation, we could reverse the unitary to teleport the post-selected state backwards in time.  However since teleportation involves measurements and classical communication which cannot be sent backwards in time we cannot simply time-reverse the usual teleportation.  

Also note that a paper with a related title, "Quantum mechanics of time travel
through post-selected teleportation" \cite{timeTravel}, creates a loop in time by post-selecting on one particular outcome of the regular (pre-selected) teleportation protocol, so that the current (pre-selected) state of a particle was sent into its past.  Here we teleport a post-selected state to a different location.

Port-Based Teleportation, introduced by Ishizaka and Hiroshima \cite{portBasedTeleportation,portBasedTeleportation2}, is one of the most useful and surprising variants of teleportation.  In normal teleportation Bob initially receives Alice's state scrambled by a unitary $\sigma_i$, and has to wait for Alice's classical message $i$ before he can unscramble it.  This means that he cannot use Alice's state in any further experiments (measurements, computations, etc) without waiting for that message (except for the rare cases where his experiment commutes with $\sigma_i$).
Port-Based teleportation instead uses $n$ entangled pairs (ports) between Alice and Bob, and teleports Alice's state unscrambled into Bob's $i^{th}$ port, where $i$ is known to Alice.  Bob can run his further experiments without waiting for Alice so long as he does it $n$ times: once on each port.  Then when he receives Alice's message he can classically select which set of experimental results to keep.  

Port-Based Teleportation has proved useful for a variety of tasks, for example it was used to make a universal programmable quantum processor \cite{portBasedTeleportation}, for instantaneous non-local quantum computation, instantaneous non-local measurements and new attacks on position-based cryptography \cite{instantaneousQuantumComputation}, for proving any communication complexity problem with a large quantum advantage gives rise to a Bell Inequality \cite{portBasedCcAndBell}, and for limitations on quantum channel discrimination \cite{portBasedChannelDiscrimination}.  It can be done either deterministically as described above, where the state is always teleported but with some error, or probabilistically where the state is either teleported perfectly or the teleportation fails with an error message.  \cite{portBasedTeleportation2}.  The choice of shared entangled state used for the teleportation and the measurement have been optimized in \cite{portBasedImprovements1,portBasedImprovements2,portBasedImprovements3, portBasedImprovements4, portBasedImprovements5}.  A multiport variation allowing teleportation of many states simultaneously into different ports was introduced in \cite{portBasedMulti1, portBasedMulti2, portBasedMulti3}.

We show how to port-based teleport a pre and post-selected state.  This can be done either deterministically or probabilistically.  We show how to use this as a quantum memory for the state.  We then show how to perform instantaneous non-local computation of pre and post-selected systems.  Instantaneous means that we perform the quantum parts of the experiments simultaneously at spatially separated locations, and then use classical communication to select the final results.

Finally, we demonstrate one use of this to significantly reduce the amount of entanglement required to show that any $\it{verification}$ measurement of spatially separated pre and post-selected systems may be performed instantaneously.  A verification measurement is one which gives the same results as a normal measurement, only it may leave the system in any output state, potentially destroying it completely.  Whether an arbitrary non-local variable of pre-selected spatially separated systems is measurable has been debated since Landau and Peierls \cite{landauPeierls} in 1931, and is important in demonstrating in a direct fashion that an entangled state is a good description of a spatially separated system in a given relativistic frame at a given time.  

After significant progress in \cite{bohrRosenfeld,aharonovAlbert1,aharonovAlbert2,aharonovAlbert3,aharonovAlbert4,aharonovAlbert5,popescuVaidman,groismanVaidman,groismanReznik}, Vaidman \cite{NonLocalMeasurementLev} showed that any non-local verification measurement can be performed instantaneously on an arbitrary pre-selected state.  The procedure used enormous amounts of entanglement, which was significantly reduced in \cite{clark_2010, instantaneousQuantumComputation, qubitInstantaneousComputation}.  It was extended to pre and post-selected systems in \cite{nonLocalPreAndPostSelectedLev}, which we improve upon here.

\section{Pre and Post-Selected Teleportation}

\label{prePostSelectRegularTeleportSection}

\begin{figure}[ht]
\includegraphics[width=8.1cm]{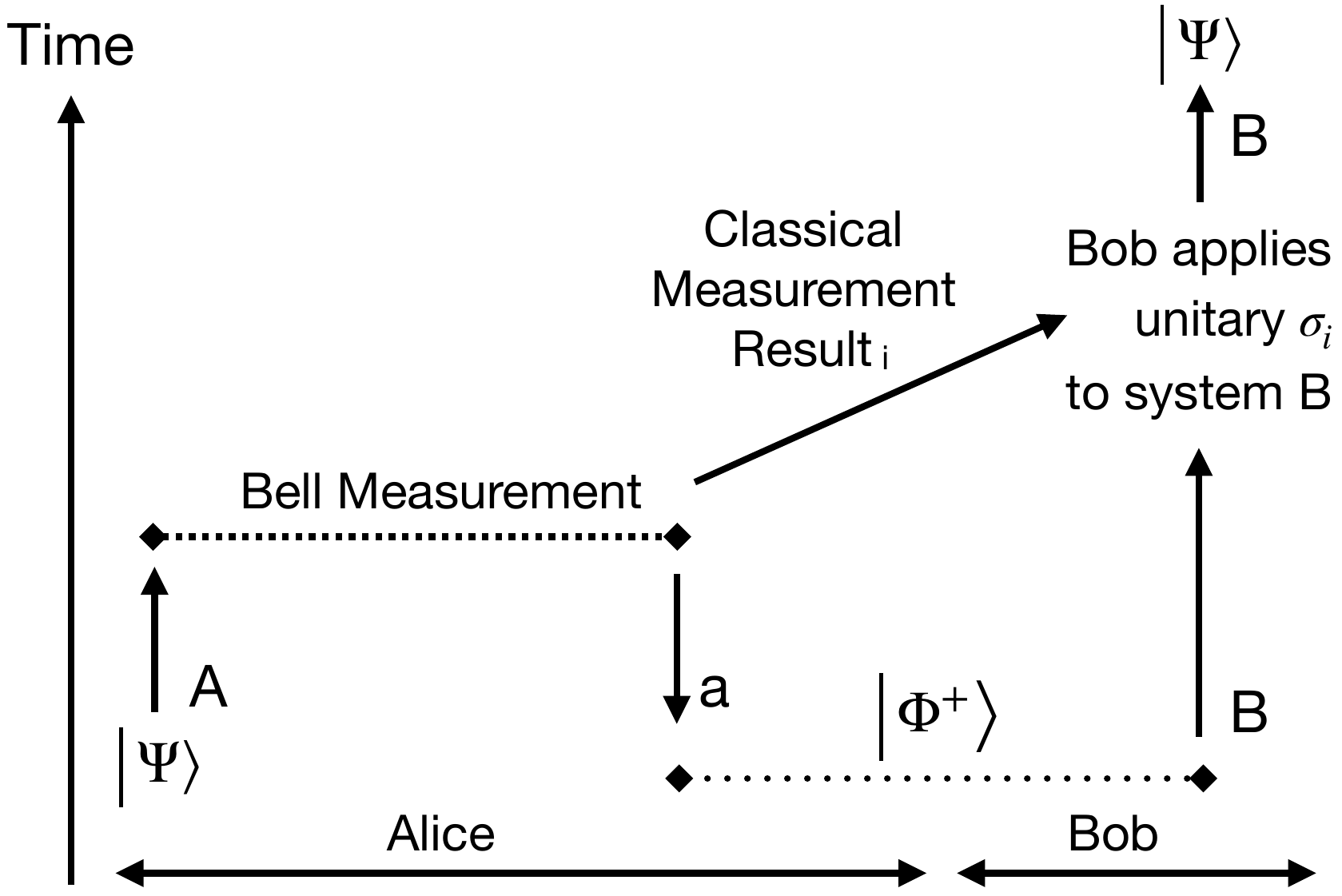}
\caption{Teleportation of a pre-selected state $\ket{\Psi}$ from $A$ to $B$.  Vertical arrows indicate the flow of the state.}
\label{Fig:PreSelect}
\end{figure} 

First we review how to teleport a pre-selected state, shown diagrammatically in Fig \ref{Fig:PreSelect}.  

For simplicity we start with a qubit which has been pre-prepared in an unknown pure state, say 
\beq
\ket{\Psi}_A = \alpha\ket{0}_A + \beta\ket{1}_A,
\eeq
where $\alpha$ and $\beta$ are complex and $|\alpha|^2 + |\beta|^2 = 1$.  Alice and Bob also pre-share a maximally entangled state 
\beq \ket{\Phi^+}_{aB} = \frac{1}{\sqrt{2}}(\ket{0}_a\ket{0}_B + \ket{1}_a\ket{1}_B). \eeq  
The joint state $\ket{\Psi}_A \ket{\Phi^+}_{aB}$ can be written as
\beq \begin{split} 
&\frac{1}{2} \{ \frac{1}{\sqrt{2}}(\ket{0}_A\ket{0}_a + \ket{1}_A\ket{1}_a)(\alpha\ket{0}_B + \beta\ket{1}_B) \\ 
&+ \frac{1}{\sqrt{2}}(\ket{0}_A\ket{0}_a - \ket{1}_A\ket{1}_a)(\alpha\ket{0}_B - \beta\ket{1}_B) \\
&+ \frac{1}{\sqrt{2}}(\ket{0}_A\ket{1}_a + \ket{1}_A\ket{0}_a)(\alpha\ket{1}_B + \beta\ket{0}_B) \\
&+ \frac{1}{\sqrt{2}}(\ket{0}_A\ket{1}_a - \ket{1}_A\ket{0}_a)(\alpha\ket{1}_B - \beta\ket{0}_B) \} \\
\end{split} \eeq
which can also be written as
\beq
\frac{1}{2} \sum_i \sigma^a_i \ket{\Phi^+}_{Aa} \sigma^B_i \ket{\Psi}_B
\eeq
where $i\in\{0,1,2,3\}$, and $\sigma_i$ is a unitary operation which flips $\ket{0}$ and $\ket{1}$ if $i\in\{2,3\}$ and applies a phase of -1 to $\ket{1}$ if $i\in\{1,3\}$.  Note that $(\sigma_i)^2 = \mathbb{1}$.

To teleport Alice measures the joint system $Aa$ in the Bell Basis $\sigma^a_i \ket{\Phi^+}_{Aa}$
and sends the measurement outcome $i$ to Bob via 2 bits of classical communication.  Bob then applies $\sigma_i$ to his system B to recreate Alice's original state $\ket{\Psi}$ in system $B$.  He can then perform whatever experiment on the state he desires, and will get the same outcomes as if he started with $\ket{\Psi}_B$.

\begin{figure}[ht]
\includegraphics[width=8cm]{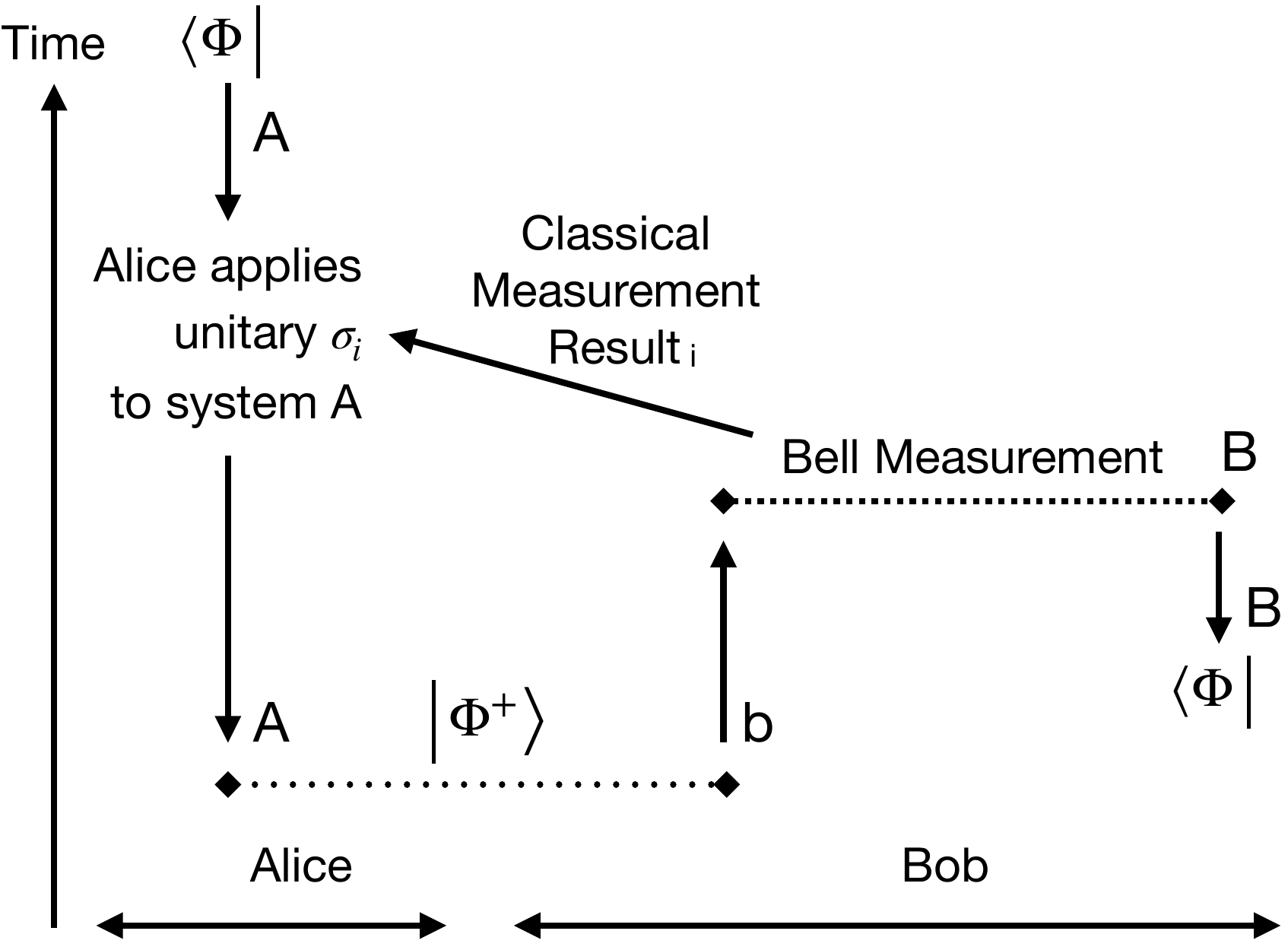}
\caption{Teleportation of a post-selected state $\bra{\Phi}$ from $A$ to $B$.  The classical communication goes from Bob to Alice, the opposite of the usual pre-selected teleportation.}
\label{Fig:PostSelect}
\end{figure}

A procedure to teleport a post-selected state $\bra{\Phi}$ should work as follows.  Alice starts with a post-selected state.  What that means is that she starts with a system in the totally uncertain pre-selected density matrix $\mathbf{1}/d$ where $d$ is the dimension of the system.  Next she does the teleportation.  Finally she does the post-selection by performing a measurement which with some probability has outcome $\Phi$, and Alice and Bob only keep the experimental results for runs with that outcome.  Bob will also have a system which starts in the pre-prepared density matrix $\mathbf{1}/d$.  He wants to be able to run experiments on his system and know that so long as the post-selection succeeds, his system will behave with the same post-selection as if he had applied it directly himself.  Further, the probability of the post-selection succeeding is the same as if he applied it himself.  This is what we mean by the teleportation succeeding with certainty.  To do this may at first appear impossible, since if Bob performs the Bell measurement from the usual teleportation on a joint system $bB$ he has no opportunity to apply $\sigma_i$ on $B$ to correct for any outcome other than $\ket{\Phi^+}$.  Our main result is a protocol to teleport with certainty, as shown in Fig \ref{Fig:PostSelect}.

Similar to the usual teleportation of pre-selected states, Alice and Bob pre-share the maximally entangled state $\ket{\Phi^+}_{Ab}$.  Bob has another system $B$ which will end up in the post-selected state.  This time Bob performs the Bell Measurement, which he does on on the joint system $bB$ after he's finished his experiments on $B$.  He then sends the classical measurement result to Alice, who performs the correction $\sigma_i$ on her system $A$.  Finally Alice applies the post-selection.  This procedure guarantees that $B$ was in the post-selected state $\bra{\Phi}$ prior to the Bell Measurement.

In the usual teleportation one can think of the movement of the state as starting with Alice, going forward in time in system $A$ until the Bell Measurement, then going backward in time (scrambled by $\sigma_i$) in system $a$ until getting to the entangled state, then going forward in time in $B$ until $\sigma_i$ is applied after which the state is successfully teleported.  In the teleportation of a post-selected state one can think of the state starting with Alice, going backwards in time in system $A$ with $\sigma_i$ applied, then backwards further until the entangled state, then forwards in time in system $b$ until the Bell Measurement which undoes $\sigma_i$, and finally backwards in time in system $B$ as the successfully teleported state. 

The teleportation actions in the two procedures are the same if we swap the roles of Alice and Bob, and hence the equations of the post-selected teleportation work in the same way as the pre-selected teleportation.  One way to check this mathematically starts by noting that the Bell Measurement with outcome $i$ is a projection onto the state $\braL{bB}{\Phi^+} \sigma^b_i$ which occurs with probability $\frac{1}{4}$. 
Working backwards in time, the post-selected teleportation does:
\beq \begin{split} 
\braL{A}{\Phi} &\rightarrow \braL{A}{\Phi} \sigma^A_i \\
              &\rightarrow \braL{A}{\Phi} \sigma^A_i \braL{bB}{\Phi^+} \sigma^b_i \\
              &\rightarrow \braL{A}{\Phi}  \braL{bB}{\Phi^+} \sigma^A_i \sigma^b_i \ket{\Phi^+}_{Ab} \\
              &= \braL{A}{\Phi} \braL{bB}{\Phi^+} \ket{\Phi^+}_{Ab} \\
              &= \frac{1}{2} \braL{B}{\Phi}
              \end{split} \eeq
as desired.  In the second line the Bell measurement acts as the preparation of a post-selected state.

This procedure is optimal in its usage of entanglement and classical communication, as proved in Appendix \ref{AppendixResources}.  The extension to higher dimensions is straightforward and follows the method in \cite{teleportation}.  Entanglement swapping and teleportation of mixed states work as usual.

\begin{figure}[ht]
\includegraphics[width=8.1cm]{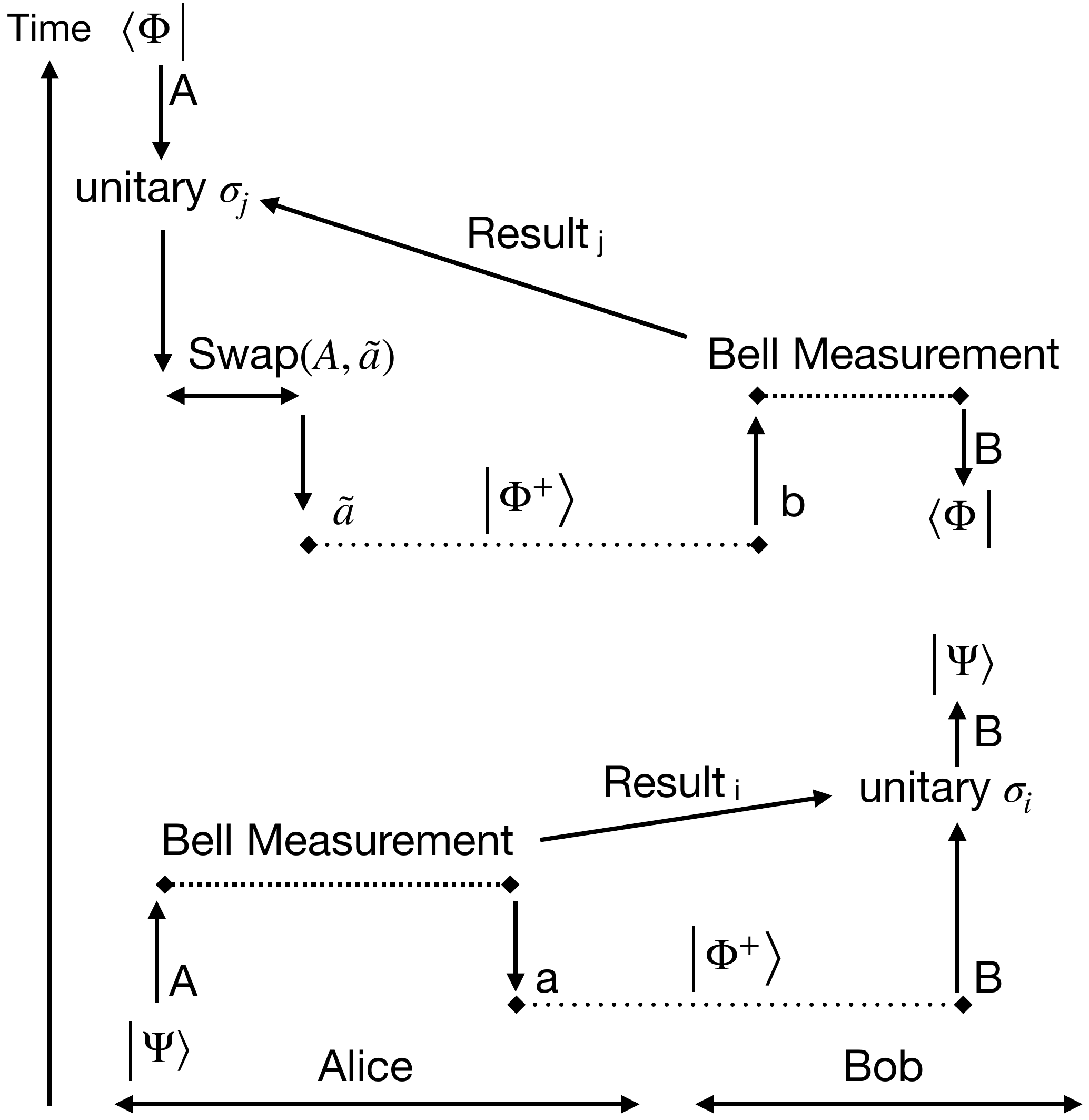}
\caption{Teleportation of a pre and post-selected state $\bra{\Phi} \ket{\Psi}$ from $A$ to $B$}
\label{Fig:PrePostSelect}
\end{figure}

Teleporting a pre and post-selected state from $A$ to $B$ is essentially a matter of combining the previous two protocols, as shown in Fig \ref{Fig:PrePostSelect}.  The only minor change is that here we used $Swap(A,\tilde{a})$ to move the entangled state $\ket{\Phi^+}_{\tilde{a}b}$ into $Ab$ for the post-selected teleportation.

This teleportation works for entangled pre and post-selected states, e.g. $\alpha \bra{\Phi_1} \ket{\Psi_1} + \beta \bra{\Phi_2} \ket{\Psi_2}$ \cite{postSelectMore}, and for mixed pre and post-selected states \cite{postSelectMixed}.  It straightforwardly generalizes to higher dimensions.  

\section{Port-Based Teleportation}

\begin{figure}[ht]
\includegraphics[width=8.1cm]{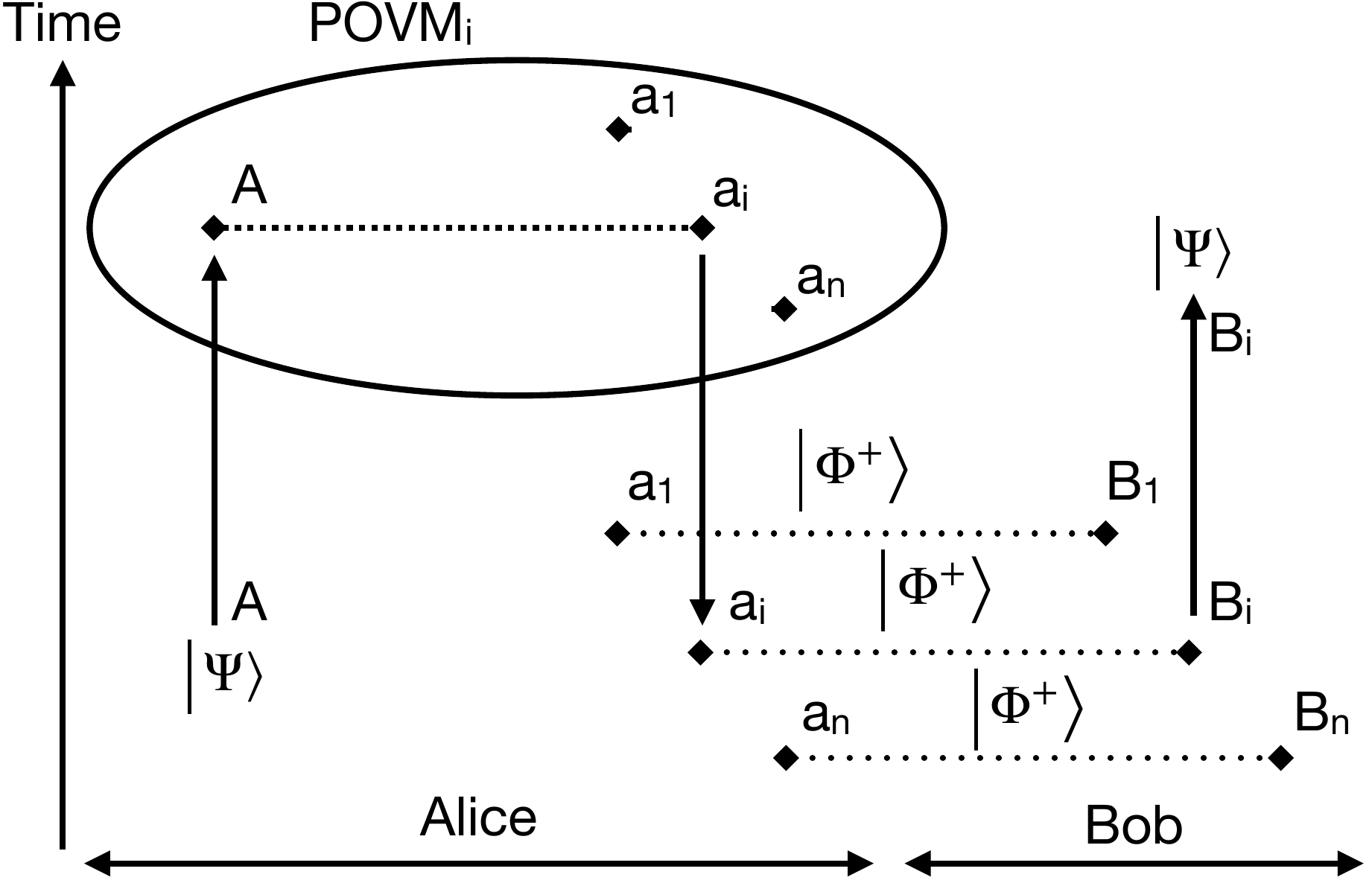}
\caption{Port-based teleportation of a pre-selected state $\ket{\Psi}$ from $A$ to $B_i$.  Outcome $i$ of the POVM moves $\ket{\Psi}$ into $a_i$ and hence $B_i$.}
\label{Fig:PortPreSelect}
\end{figure}

Here we show how to port-based teleport a post-selected state, and then a pre and post-selected state.  We then use this as a quantum memory for such states.  We start by reviewing port-based teleportation of a pre-selected state, as shown in Fig \ref{Fig:PortPreSelect}.  

Alice and Bob pre-share $n$ entangled states $\ket{\Phi^+}_{a_iB_i}$, with $i=1..n$.  Alice performs a POVM on the joint system $Aa_1..a_n$.  The outcome $i$ tells her that $\ket{\Psi}_A$ has been teleported into $\ket{\Psi}_{B_i}$.  There are two main variations of the port-based teleportation.  In the deterministic version, the teleported state is unmodified, except for some noise which can be made arbitrarily small with sufficiently large $n$.  In the probabilistic version, the teleportation either succeeds perfectly or with an arbitrarily small probability Alice's POVM gives a failure outcome, $0$.  Unlike usual teleportation there are several POVMs and initial entangled states which could be used, and the most efficient procedures do not use the maximally entangled state.  However schematically the procedure is always the same.  In all cases Bob can perform his further experiments, $n$ times in parallel, without waiting for Alice to perform the quantum part of the teleportation.  After the quantum part is finished Alice can send $i$ as a classical message to Bob to tell him which of his results to keep.  This will take some time as it's limited by the speed of light.

\begin{figure}[ht]
\includegraphics[width=8.1cm]{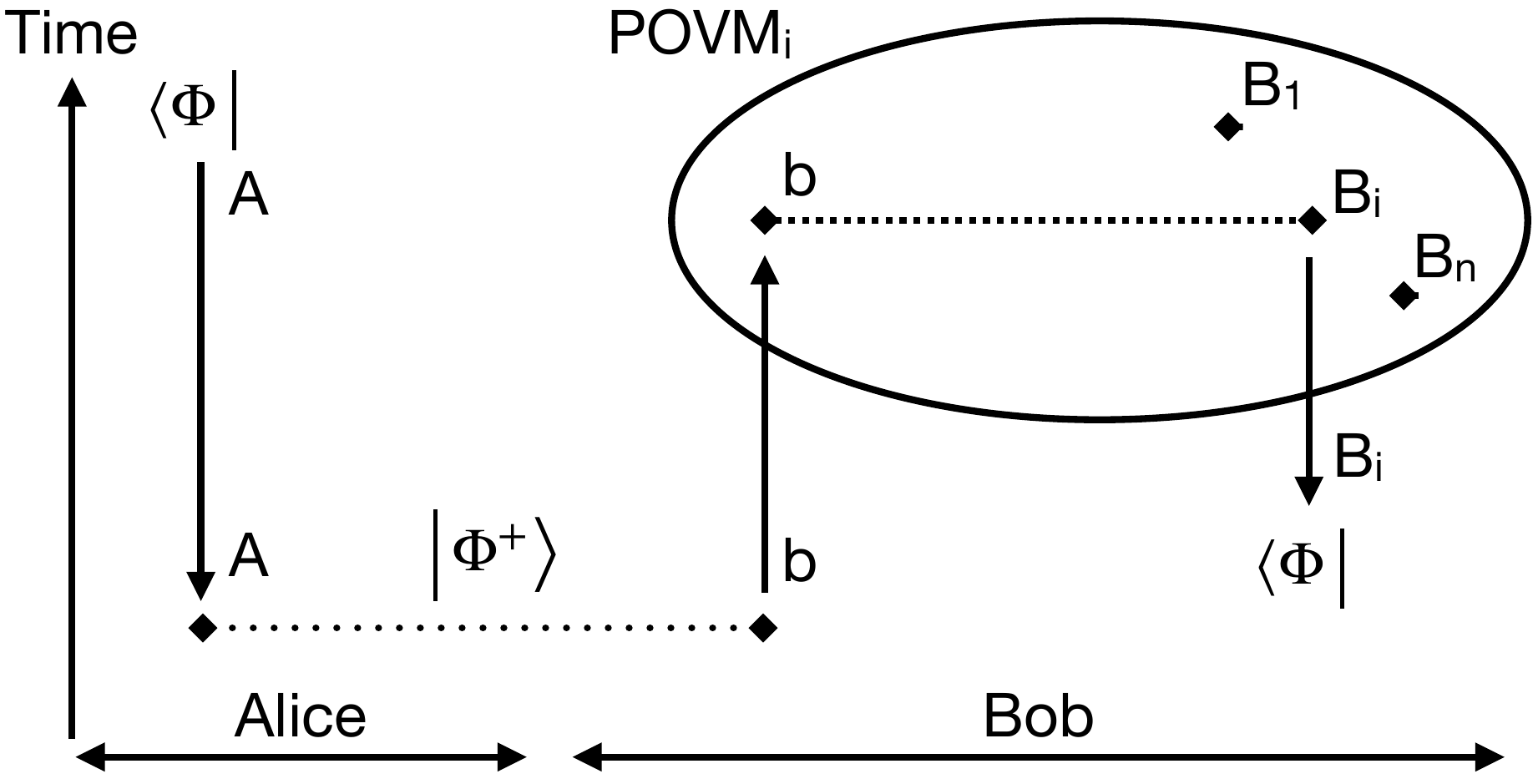}
\caption{Port-based teleportation of a post-selected state $\bra{\Phi}$ from $A$ to $B_i$.}
\label{Fig:PortPostSelect}
\end{figure}

In the post-selected teleportation described in section \ref{prePostSelectRegularTeleportSection}, Bob performs his experiments at the start, but Alice cannot apply the post selection without waiting for Bob's classical message.  Post-selected port-based teleportation (Fig \ref{Fig:PortPostSelect}) instead allows Alice to post-select without waiting for Bob.

%Note: I've put this next figure here higher than it's referenced in the text in order to get it to display in the right place.  If you change the text, please check the position of the figures again and move if necessary!
\begin{figure*}[ht]
\includegraphics[width=14cm]{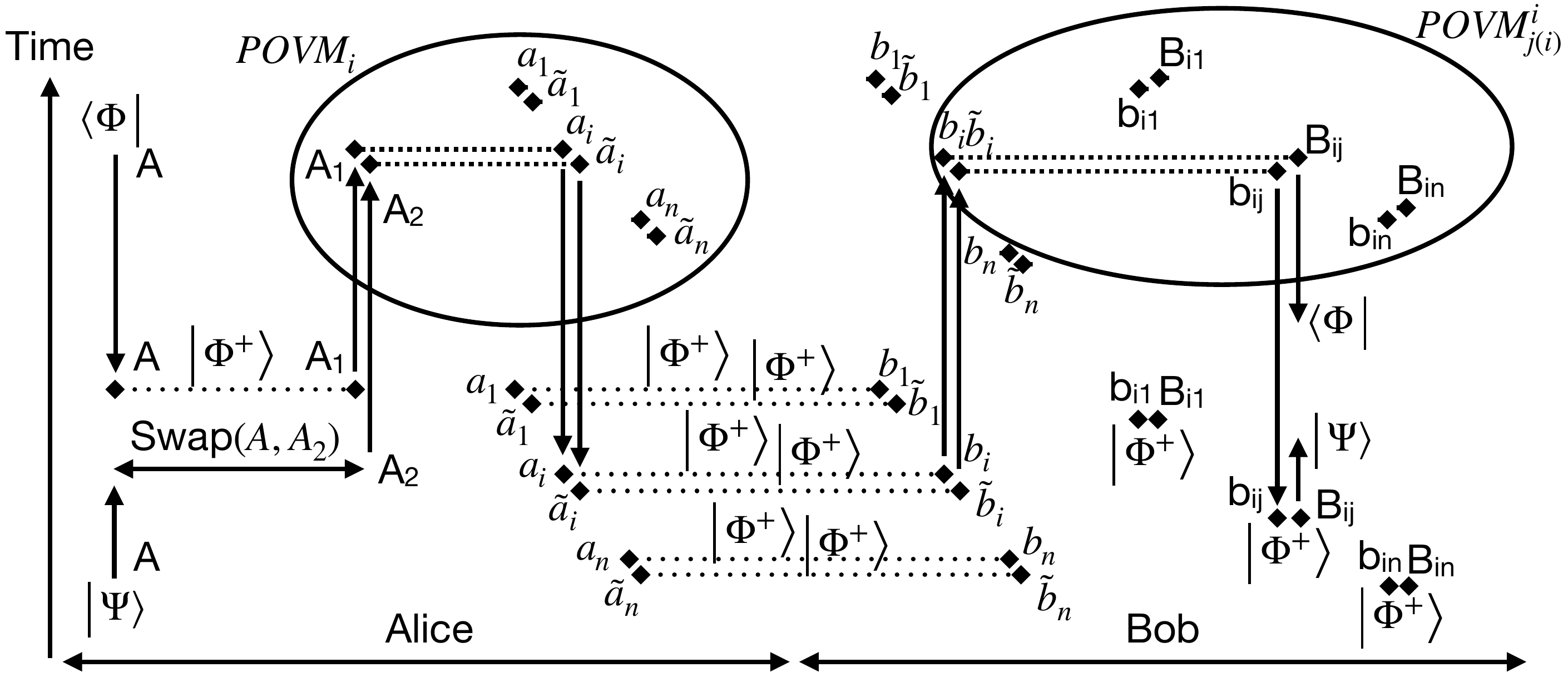}
\caption{Port-based teleportation of a pre and post-selected state $\bra{\Phi} \ket{\Psi}$ from $A$ to port $B_{ij}$.  Alice's POVM has outcome $i$.  Bob performs $n$ POVMs: for compactness we only draw the $i^{th}$, $POVM^i$ with outcome $j(i)$.  Each port-based channel transmits the state in two systems together, e.g. $A_1$ and $A_2$, so has dimension $d^2$ where $d$ is the dimension of the original system $A$.}
\label{Fig:PortPrePostSelect}
\end{figure*}

Suppose the system $A$ will be post-selected into $\braL{A}{\Phi}$.  Alice and Bob have a pre-shared entangled state $\ket{\Phi^+}_{Ab}$.  Bob has systems $B_i$ where $i=1..n$ (the $n$ ports), and in each of them runs his experiment.  He then applies the usual port-based teleportation POVM on the joint system $bB_1B_2..B_n$.  POVM outcome $i$ tells him that the post-selected state was teleported into system $B_i$ (up to the usual port-based error in the deterministic version, and perfectly unless we get the failure outcome $0$ in the probabilistic version), and so his experiment was run with that post-selection.  Unlike most teleportation protocols, it seems as though Bob does not need any classical communication from Alice.  However he does need one message: that the post selection succeeded.  Finally, note that the POVM is a method for transforming a pre-selected state into a post-selected state in channel $i$ without any scrambling, which we shall use as a primitive in later protocols.

Next we do port-based teleportation on a pre and post-selected state $\braL{A}{\Phi} \ket{\Psi}_A$, as shown in Fig \ref{Fig:PortPrePostSelect}.  We cannot simply do pre-selected teleportation followed by post-selected teleportation, as the pre and post-selections could end up in different ports.  
We thus need to somehow move the pre and post-selected states together.  For this we will use pre-selected port-based teleportation on a $d^2$ dimensional system, where $d$ is the dimension of the Hilbert space of $A$.  In stages:

\begin{enumerate}
\item Alice converts her pre and post-selected state $\braL{A}{\Phi} \ket{\Psi}_A$ into a pre-selected state $\ket{\Phi^*}_{A_1} \ket{\Psi}_{A_2}$ by swapping $\ket{\Psi}_A$ into $\ket{\Psi}_{A_2}$ and then preparing $\ket{\Phi^+}_{AA_1}$.  Here $``^*"$ means complex conjugation in the $\ket{0}/\ket{1}$ basis.
\item Alice pre-selected port-based teleports the joint system $\ket{\Phi^*}_{A_1} \ket{\Psi}_{A_2}$ to Bob, and it arrives in port $i$ (the outcome of her POVM) as  $\ket{\Phi^*}_{b_i} \ket{\Psi}_{\tilde{b}_i}$.
\item For each $i$ Bob performs the port-based $POVM^i$, with outcome $j(i)$, to convert this into the post-selected state $\braL{B_{i,j}}{\Phi} \braL{b_{i,j}}{\Psi^*}$ in port $(i,j)$. 
\item Bob transforms each $(i,j)$ into the original pre and post-selected state $\braL{B_{i,j}}{\Phi} \ket{\Psi}_{B_{i,j}}$ by pre-preparing the state $\ket{\Phi^+}_{B_{i,j}b_{i,j}}$, and leaving system $b_{i,j}$ otherwise untouched.
\item Alice sends $i$ to Bob using classical communication, and Bob picks out the results from port $(i,j)$.
\end{enumerate} 

Thus we have teleported a pre and post-selected state from $A$ into port $B_{i,j}$ with $i$ known to Alice and $j(i)$ known to Bob.  Neither Alice nor Bob have to wait for one another before completing the quantum part of their experiments, including the post-selection.  As before, these methods work for entangled and mixed pre and post-selected states.  As the protocols are built upon the usual port-based teleportation POVM and pre-shared entangled state, we can use whichever port-based variant we like, deterministic or probabilistic, or multi-port.

%Note: I've put this next figure here higher than it's referenced in the text in order to get it to display in the right place.  If you change the text, please check the position of the figures again and move if necessary!
\begin{figure*}[ht]
\includegraphics[width=14cm]{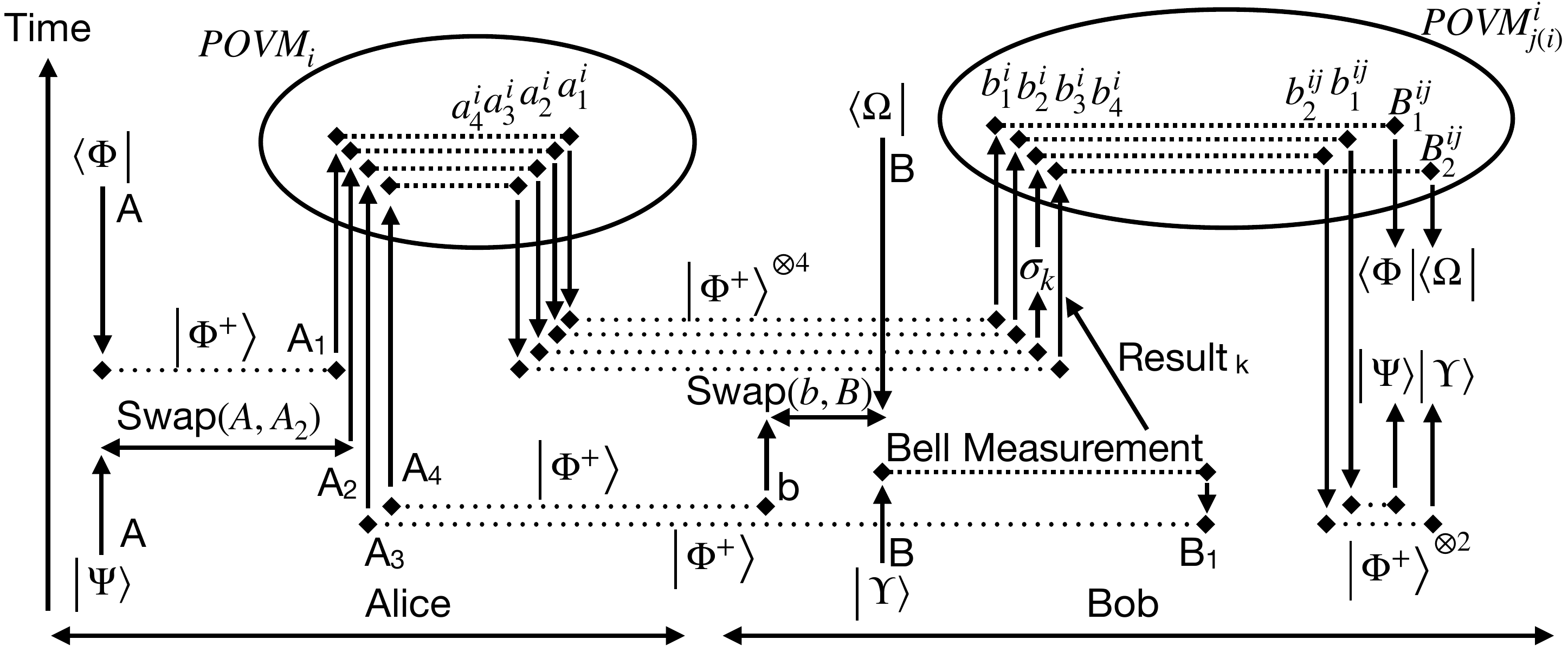}
\caption{Instantaneous Computation/Measurement on a pre and post-selected state $\bra{\Phi} \ket{\Psi}$ for Alice and $\bra{\Omega} \ket{\Upsilon}$ for Bob.  Alice and Bob move their initial states into port $B_{ij}$ on Bob's side, with $i$ known to Alice and $j(i)$ to Bob.  Bob can perform any computation or measurement on the joint system, doing it $n^2$ times (once on each port). We have omitted the other ports and POVMs from the picture for compactness.}
\label{Fig:InstantaneousMeasure}
\end{figure*}

One use of this protocol is as a quantum memory for (pre and) post-selected states.  We may be given a state, and wish to store it for later use with a quantum channel which we can use multiple times.  This is in a sense a time-reversed application corresponding to the storage of a quantum channel for usage later on a pre-selected quantum state, for which the optimal protocol is described in \cite{storeUnitary}.  We can store the pre and post-selected state by following Alice's steps in the port-based teleportation protocol just described in Fig \ref{Fig:PortPrePostSelect}.  Later on, when we have the channel ready to use, we perform Bob's steps of the teleportation protocol, applying the channel in every single port $B_{i,j}$.  Finally we read out the results from port $(i,j)$.  Depending upon whether we use deterministic or probabilistic port-based teleportation, this will have the same noise or probability of failure as our teleportation protocol, which can be made as small as desired by increasing the number of ports.  We do not claim this protocol is optimal, and leave such an investigation for future work. 

\section{Instantaneous Non-Local Computation and Measurement}

We show in Fig \ref{Fig:InstantaneousMeasure} how to perform an arbitrary $\it{instantaneous}$ quantum computation or measurement on two spatially separated pre and post-selected systems, generalizing the result for pre-selected systems in \cite{instantaneousQuantumComputation}. ``Instantaneous" means that we perform the quantum parts of the experiments simultaneously at spatially separated locations, and then use classical communication to select the final results.  Whilst the quantum part is instantaneous, the classical communication takes time.  The protocol can be extended to any number of parties.  

Suppose Alice and Bob wish to perform a joint computation on systems $A$ and $B$ which are in a pre and post-selected state $\braL{A}{\Phi} \ket{\Psi}_A$ for Alice and $\braL{B}{\Omega} \ket{\Upsilon}_B$ for Bob.  They could try to use a port-based teleportation from Bob's system $B$ to Alice's ancilla system $a_i$ to move the whole state together on Alice's side to allow her to perform the computation.  However this puts the state into the system $Aa_i$ where $i$ is known only to Bob, and Alice cannot make an arbitrary bi-partite measurement simultaneously on $Aa_0$, $Aa_1$ etc as they will disturb each another on $A$.  Instead they could try (non-port based) teleporting Bob's state to Alice (without initially doing any unscrambling as that requires waiting for classical communication), and then port-based teleporting the joint state back to Bob.  However both $\bra{\Phi}$ and $\ket{\Psi}$ are scrambled, and Bob only knows how to unscramble $\ket{\Psi}$.  Instead they skip the step causing the scrambling of $\ket{\Psi}$, which gives the successful protocol shown in Fig \ref{Fig:InstantaneousMeasure}.  Alice and Bob
\begin{enumerate}
\item Start by bringing Alice and Bob's systems together on Alice's side as a $4$ system pre-selection $\ket{\Phi^*}_{A_1}  \ket{\Psi}_{A_2} \sigma^{A_3}_k \ket{\Upsilon}_{A_3} \ket{\Omega^*}_{A_4}$, where $\ket{\Upsilon}$ is scrambled by $\sigma_k$ known to Bob.
\item Alice port-based teleports this joint state to Bob as a single $d^4$ dimensional system into port $i$, $\ket{\Phi^*}_{b^i_1} \ket{\Psi}_{b^i_2} \sigma^{b^i_3}_k \ket{\Upsilon}_{b^i_3} \ket{\Omega^*}_{b^i_4}$, where $i$ is known to Alice.  
\item Bob applies $\sigma_k$ on $b^i_3$ for each $i$ to unscramble $\ket{\Upsilon}_{b^i_3}$.  
\item For each $i$ Bob performs the port-based $POVM^i$, with outcome $j(i)$, to convert this into the post-selected state $\braL{B^{ij}_1}{\Phi} \braL{b^{ij}_1}{\Psi^*} \braL{b^{ij}_2}{\Upsilon^*} \braL{B^{ij}_2}{\Omega}$ in port $(i,j)$. 
\item Bob transforms each $(i,j)$ into the original pre and post-selected state $\braL{B^{i,j}_1}{\Phi} \ket{\Psi}_{B^{i,j}_1} \braL{B^{i,j}_2}{\Omega} \ket{\Upsilon}_{B^{i,j}_2}$ by pre-preparing the state $\ket{\Phi^+}_{B^{i,j}_1 b^{i,j}_1} \ket{\Phi^+}_{B^{i,j}_2 b^{i,j}_2}$ and leaving systems $b^{i,j}_1$ and $b^{i,j}_2$ otherwise untouched. 
\end{enumerate}

Bob can perform whatever computation or measurement,  von Neumann or POVM, he wants on the joint pre and post-selected system $B^{ij}_1 B^{ij}_2$, repeating it for all ports $(i,j)$.  If he wishes he can use pre and post-selected port-based teleportation to send all the systems $B^{ij}_1$ back to Alice, using the multi-port variation \cite{portBasedMulti1, portBasedMulti2, portBasedMulti3} for efficiency.  At the end Alice knows $i$ and Bob knows $j(i)$, so they can pick the right results via classical communication.  That completes the protocol, which again works on arbitrary initial entangled states, pure or mixed. 

This port-based protocol is exponentially more efficient in terms of entanglement than the previously known instantaneous measurement protocol for pre and post-selected states \cite{nonLocalPreAndPostSelectedLev}.  This exponential gap was shown in \cite{instantaneousQuantumComputation} for the pre-selected case.  The ebits of entanglement used to instantaneously measure a system of $m$ qubits for Alice (i.e. dimension $d = 2^m$) and $m$ for Bob, whilst making the probability of failure less than $\epsilon$, scales with $m$ as shown in table \ref{table:entanglementUsage}.  Our protocol is more efficient still for POVMs as it uses the same entanglement as a von Neumann measurement, whereas the previous protocol requires writing the POVM as a von Neumann measurement using an ancilla, then teleporting the ancilla and system to be measured together back and forth as one higher dimensional system.
\begin{table}[ht]
\centering
\tabcolsep=0.11cm
\begin{tabular}{ c | c | c } 
 & \multicolumn{2}{c}{Protocol} \\
 \cline{2-3}
Type     & Original & Port-Based \\ 
 \hline 
 Pre            & $\mathlarger{\mathlarger{4m 2^{\ln(1/\epsilon) 4m 2^{4m} - 6m}}}$        & $\mathlarger{\mathlarger{(2m/\epsilon) 2^{4m}}}$ \\ 
 Pre\&Post   & $\mathlarger{\mathlarger{8 m 2^{\ln(1/\epsilon) 8m 2^{8m} - 14m}}}$        & $\mathlarger{\mathlarger{(4m/\epsilon) 2^{8m}}}$ \\ 
\end{tabular}
\caption{Approximate entanglement (ebits) used to perform an instantaneous measurement on $m$ qubits for Alice and $m$ for Bob with $\epsilon$ probability of success.  The port-based numbers are an upper bound.  The port-based scheme is exponentially better in $m$.}
\label{table:entanglementUsage}
\end{table}

To calculate that upper bound of the entanglement usage for the port-based instantaneous measurement for pre-selected systems, we start from the optimal probability of success of port-based teleportation, given by theorem 4 in \cite{portBasedImprovements2}, as $p = 1 - (d^2-1)/(n + d^2 - 1)$, where $n$ is the number of ports.  The measurement protocol first teleports $m$ qubits from Bob to Alice, using $m$ ebits.  Then it port-based teleports all $2m$ qubits back from Alice to Bob as a single $d = 2^{2m}$ dimensional system in $n$ ports, using at most $2m n$ ebits (the optimal port-based teleportation uses a non-maximally entangled state whose entanglement will be upper bounded by this).  Setting $p = 1 - \epsilon$, $\epsilon$ small so we can ignore the entanglement usage of the original teleportation, and ignoring terms which are small compared to $2^{4m}$, we have an upper bound on the entanglement usage of $(2m/\epsilon) 2^{4m}$.  The entanglement usage of measuring $m$ pre and post-selected qubits is similar, with $m$ replaced by $2m$, as each pre and post-selected qubit is mapped into $2$ pre-selected qubits before being port-based teleported.

To calculate the entanglement usage of the original (Vaidman) protocol to measure $m$ pre-selected qubits for Alice and $m$ for Bob, we follow section 1.3 of \cite{instantaneousQuantumComputation}.  The probability of success in each round of the protocol is $2^{-4m}$ (except for the first round, which is $2^{-2m}$).  Then the number of rounds, $r$, needed to make the probability of failure equal to a given $\epsilon$ is roughly $\ln(1/\epsilon) 2^{4m}$, where $\ln$ is the natural logarithm.  The entanglement usage of round $r$ (after the first couple of rounds) is roughly $4m 2^{4m(r-2) + 2m}$.  Putting these together gives the result in table \ref{table:entanglementUsage}.  The entanglement usage of the original protocol for pre and post-selected qubits \cite{nonLocalPreAndPostSelectedLev} was calculated along similar lines.

\section{Conclusion}

We have shown how to teleport pre and post-selected states with certainty.  We have also shown how to teleport them port-based.  We then showed how to perform instantaneous non-local quantum computation on a pre and post-selected state.  This allows us to store a post-selected state in a quantum memory for later usage, and to perform the quantum part of any non-local verification measurement instantaneously using exponentially less entanglement than the previous protocol.  We believe that these protocols will prove useful for other quantum information processing tasks.

\section{Acknowledgements}
We thank Sandu Popescu for helpful discussions and advice.  We are supported by the ERC Advanced Grant FLQuant.

\medskip

\textit{Note Added} After completing this work we became aware of an earlier paper by Vaidman \cite{backwardEvolvingStates} which described post-selected teleportation, similar to our section \ref{prePostSelectRegularTeleportSection}.  Our paper goes further, extending the results to post-selected {\it port}-based teleportation, and using it as a quantum memory for post-selected states, instantaneous computation, and instantaneous measurement.    

\bibliographystyle{quantum}
\bibliography{PostTeleport}

\appendix
\section{Optimality of Resources}
\label{AppendixResources}

Here we prove that the post-selected teleportation described in Fig \ref{Fig:PostSelect} is optimal in terms of entanglement and classical communication resource.

First to prove that a maximally entangled pair $\ket{\Phi^+}_{AB}$ is necessary, we show how to use the post-selected teleportation to create a maximally entangled pair $\ket{\Phi^+}_{AB}$ between Alice and Bob.  As entanglement cannot be created by local operations and classical communication, it must have been there initially.  
Suppose we start with an entangled pair on Alice's side $\ket{\Phi^+}_{Aa_1}$, another on Bob's side $\ket{\Phi^+}_{bB}$ and teleport system $a_2$ of the post-selected state $\braL{a_1a_2}{\Phi^+}$ to system $b$.  This gives us 
\beq
\braL{a_1b}{\Phi^+} \ket{\Phi^+}_{Aa_1} \ket{\Phi^+}_{bB} = \ket{\Phi^+}_{AB}
\eeq

i.e. the same entanglement as we used in the teleportation.  The post-selection may not always succeed, so we apply it by performing the Bell Measurement with outcome $j$ on $a_1a_2$, and teleporting whichever Bell state is the outcome of the measurement.  Alice can then apply a local rotation $\sigma_j$ to turn this into $\ket{\Phi^+}_{AB}$, completing the proof.   

To prove the teleportation is optimal in using two bits of classical communication from Alice to Bob, we shall show Bob can communicate two bits to Bob using Super Dense Coding \cite{superDenseCoding}.  We start with an an entangled pair on Alice's side $\ket{\Phi^+}_{a_1a_2}$, and an entangled pair between Alice and Bob $\ket{\Phi^+}_{AB}$.  Bob can encode his message $i$ by applying $\sigma^B_i$ to system $B$.  They teleport Alice's system $a_3$ in the post-selection $\braL{a_2a_3}{\Phi^+}$ to system $B$.  After the teleportation Alice will have 
\beq
\braL{a_2B}{\Phi^+} \ket{\Phi^+}_{a_1a_2} \sigma^B_i \ket{\Phi^+}_{AB} = \sigma^{a_1}_i \ket{\Phi^+}_{Aa_1}.
\eeq
She can perform a Bell measurement on $Aa_1$ to read the message.  The post-selection may not always succeed, so we perform it by applying the Bell measurement on ${a_2a_3}$.  Getting different outcomes of the Bell measurement on ${a_2a_3}$ simply permutes the outcomes of the Bell measurement on $Aa_1$ in a deterministic way, so Alice will always be able to read the message and receive 2 bits from Bob. 

\end{document}